\documentclass[12pt,draftclsnofoot, onecolumn]{IEEEtran}
\usepackage{bbm}
\usepackage{amsfonts,longtable}
\usepackage[normalem]{ulem}
\usepackage{mathrsfs}
\usepackage{graphicx,cite,epsfig,amssymb,amsmath,color}

\begin{document}

\title{Non-Orthogonal Multiple Access For Cooperative Communications: Challenges, Opportunities, And Trends}

\author{\normalsize
Dehuan~Wan, Miaowen~Wen, Fei~Ji, Hua~Yu, and Fangjiong~Chen\\
\vspace{1cm}
\thanks{This work was supported in part by the National Natural Science Foundation of China under Grant 61431005, Grant 61771202, Grant 61671211, Grant U1701265, and Grant 61501190, and in part by the Guangdong provincial research project under Grant 2016A030308006.}
\thanks{All authors are with the School of Electronic and Information Engineering, South China University of Technology, Guangzhou 510641, China (e-mail: eemwwen@scut.edu.cn).}
}

\maketitle
\begin{abstract}
Non-orthogonal multiple access (NOMA) is a promising radio access technique for next-generation wireless networks. In this article, we investigate the NOMA-based cooperative relay network. We begin with an introduction of the existing relay-assisted NOMA systems by classifying them into three categories: uplink, downlink, and composite architectures. Then, we discuss their principles and key features, and provide a comprehensive comparison from the perspective of spectral efficiency, energy efficiency, and total transmit power. A novel strategy termed hybrid power allocation is further discussed for the composite architecture, which can reduce the computational complexity and signaling overhead at the expense of marginal sum rate degradation. Finally, major challenges, opportunities, and future research trends for the design of NOMA-based cooperative relay systems with other techniques are also highlighted to provide insights for researchers in this field.
\end{abstract}

\date{\today}
\renewcommand{\baselinestretch}{1.2}
\thispagestyle{empty} \maketitle \thispagestyle{empty}

\newpage

\IEEEpeerreviewmaketitle

\section{Introduction}
The rapid development of wireless transmission and mobile communications techniques has lead to an explosive increase in the data traffic of wireless networks. In order to support the tremendous demands on data traffic, the non-orthogonal multiple access (NOMA) technique, which can be realized in the power domain, code domain, or other domains\cite{SNOK_CST_TUTORIALS2016, SBQSSK_5G_networks_TVT_2017}, is widely recognized as a key technology for the fifth generation (5G) mobile communications systems\cite{LBYSCZ_NOMA5G_future_CM2015, ZYJQMCH_5G_networks_CM_2017}.
\par
Recently, significant research efforts have been dedicated to applying NOMA techniques to various scenarios, motivated by the following benefits\cite{LBYSCZ_NOMA5G_future_CM2015}:
\begin{itemize}
\item Higher spectrum efficiency and cell-edge throughput: NOMA can serve an arbitrary number of users in each resource block by superimposing all users' signals, and according to the near-far effect, more power can be allocated to the nodes with poor channel quality to improve the system throughput.
\item Good backward compatibility with other techniques: NOMA can be easily applied on top of orthogonal frequency division multiple access (OFDMA) for downlink and single-carrier frequency-division multiple access (SC-FDMA) for uplink, and can be easily combined with massive multiple-input multiple-output (MIMO) and millimeter-wave (mmWave) techniques to further support better system performance.
\end{itemize}
Among those, the application of NOMA to cooperative relaying scenarios is becoming popular.  By now, a class of dual-hop relay-aided NOMA systems has been developed, where no direct link exists and the relay adopts either decode and forward (DF) or amplify and forward (AF) protocol. For example, the cooperative X network\cite{MMS_NOMAXrelay_CL2017} and the diamond network\cite{WMYYFF_NOMA_ICSI} considering NOMA and the DF protocol have been constructed, where the decoding of the signals at the relay and user must follow the same manner, or the achievable sum rate will deteriorate dramatically. On the other hand, the study with the AF protocol has been carried out in\cite{DMFYY_NOMAmultipleantennarelaying_TCOM18}. Besides, a downlink communication system would become a cooperative relay network if the cell-centre users can act as relay nodes for cell-edge users via successive interference cancellation (SIC)\cite{ZMH_CooperativeNOMA_CL1513}. The superiority of NOMA over OMA in terms of achievable sum rate is highly dependent on the system's asymmetry. Therefore, configuring or selecting the relays to fully exploit the near-far effect is a goal to achieve a better system performance with NOMA. However, how to realize effective configuration and optimization for such networks is still an open issue.
\par
In this article, we develop a unified architecture for the power-domain NOMA cooperative relay network through a sophisticated combination of three basic communication structures. The system settings, decoding orders, system asymmetries, power allocation, and performance are discussed. Besides, a novel hybrid power allocation strategy is proposed for the composite architecture. The major challenges, opportunities, and future research trends are also discussed.
\par
The rest of this article is organized as follows. Section II first illustrates the basic communication structures and then presents the categories of cooperative relay networks accordingly. Section III briefly reviews the features of NOMA in uplink and downlink transmissions. In Section IV, we discuss the characteristics of system settings, decoding strategies, system asymmetries, and power allocation schemes for the cooperative relay systems. Besides, performance comparisons among the existing cooperative relay architectures are presented. Section V proposes an effective hybrid power allocation strategy for NOMA cooperative relay systems. In Section VI, the opportunities, challenges, and research trends are highlighted. Finally, Section VII concludes this article.

\section{Structures of Relay-Aided Wireless Networks}
\begin{figure}[p]
\centering
\includegraphics[width=5.0in]{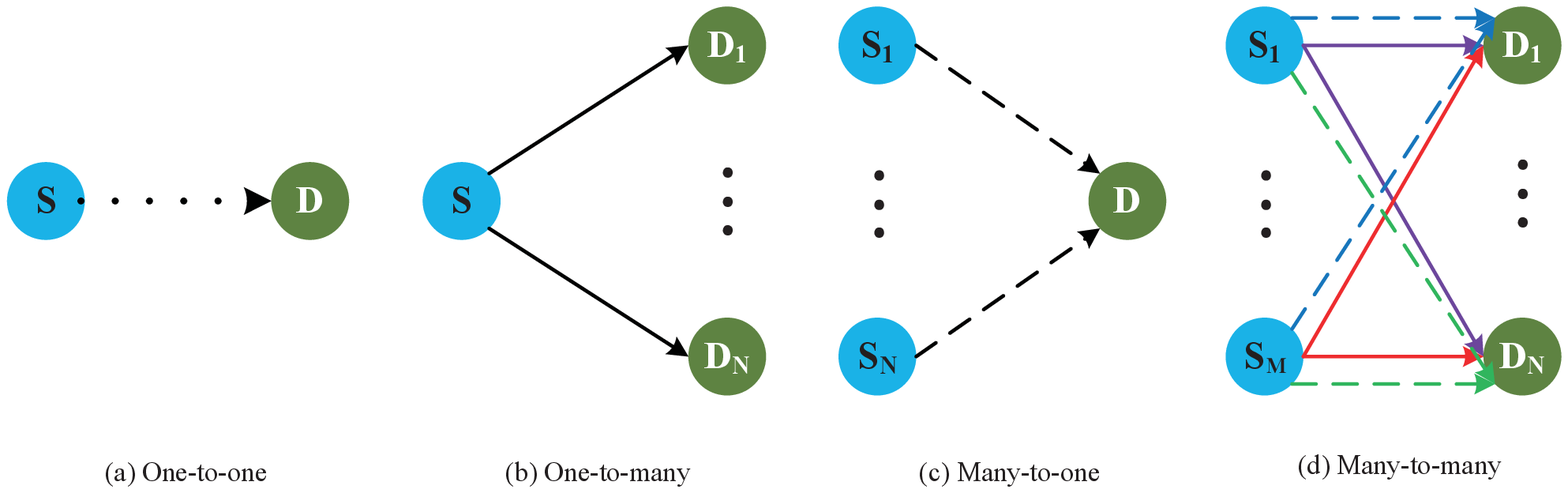}
\caption{\small Basic communication modes.}
\label{fig_1}
\end{figure}

There are three basic modes for communications between source S and destination D, namely the one-to-one, one-to-many, and many-to-one modes, as shown in Figs. \ref{fig_1}(a)-(c). Actually, there exists another many-to-many mode, as shown in Fig. \ref{fig_1}(d). However, such a structure can be constructed by the first three basic modes. For brevity, the many-to-many mode will not be discussed in this article.
\par
\begin{figure}[p]
\centering
\includegraphics[width=5.4in]{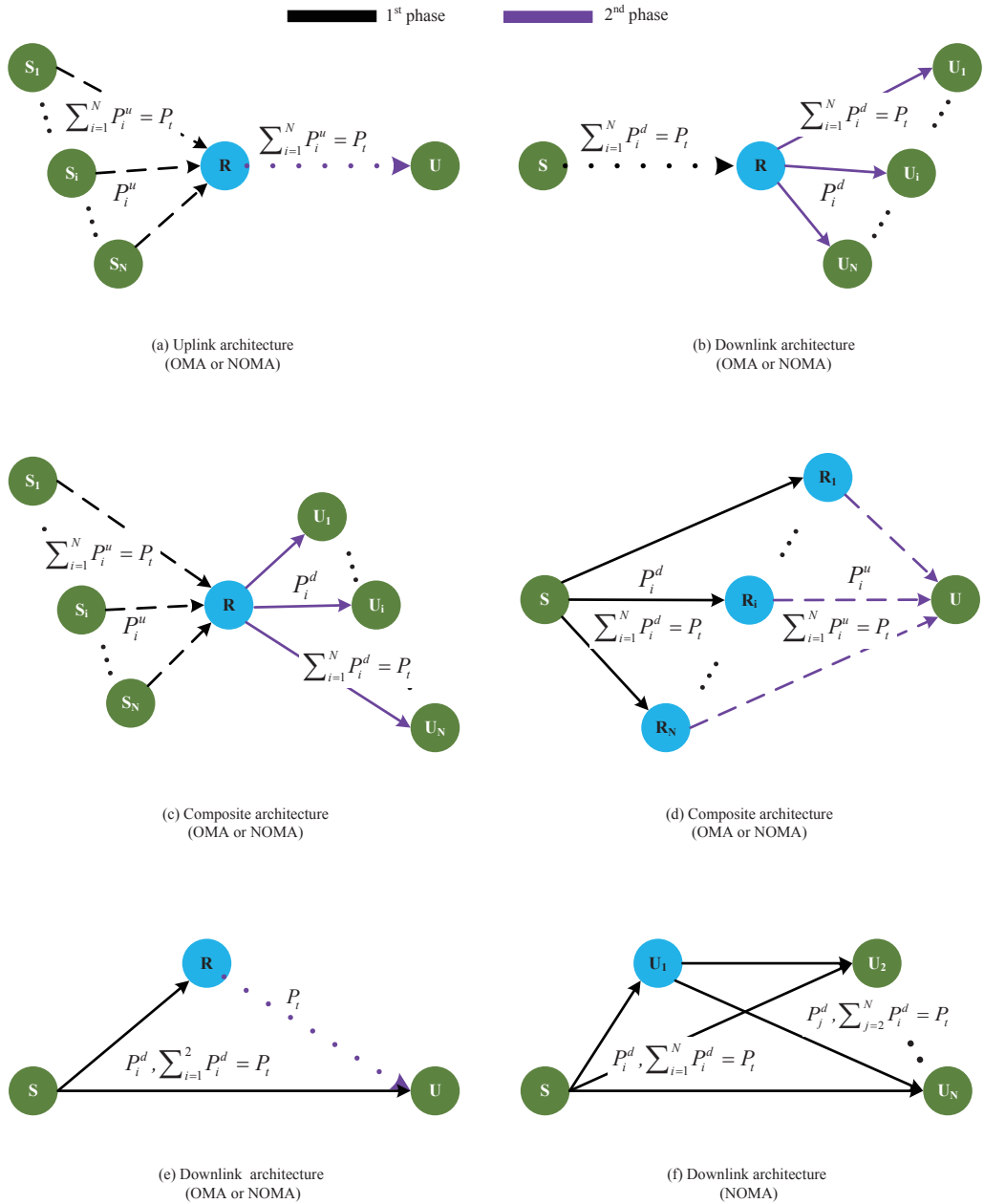}
\caption{\small Architectures of the cooperative relay systems for OMA and NOMA, where S, R, and U denote the source (acting as a transmitter), relay (acting as a transmitter and a receiver), and user (acting as a receiver in OMA, but can also act as a transmitter in NOMA), respectively, $P_t$ stands for the total power during each transmission, and the superscript $u$ or $d$ denotes the uplink or downlink transmission.}
\label{fig_2}
\end{figure}

According to the basic communication modes, the cooperative relay networks, which comprise the sources S, relays R, and users U, can be simply divided into three categories, namely the uplink (many-to-one structure), the downlink (one-to-many structure), and the composite (both uplink and downlink are comprised) architectures, as shown in Figs. \ref{fig_2}(a)-(d). Since the one-to-one mode can be subordinated to either the uplink or downlink transmission, the cooperative relay system constructed by either the uplink or downlink architecture always includes a one-to-one mode. For the case that the direct links between the source and user nodes exist, this article only focuses on the classical three-node relay-aided wireless networks, as shown in Fig. \ref{fig_2}(e). On the other hand, due to the orthogonality, the receiver in OMA can only decode the data by applying the single-user detector, such that the user node cannot act as a relay node to improve the system performance. However, for NOMA, since SIC can realize multi-user detection, the cell-centre user can act as a relay to improve the reception reliability for other users with poor connections, as shown in Fig. \ref{fig_2}(f).

\section{Features of NOMA in Uplink and Downlink}
For ease of demonstration, the channel between the base station and $i$-th user $U_i$ is denoted by $\emph{h}_{i} \sim \mathcal {CN}(0,\sigma^2_{i})$ and we measure the channel condition for $U_i$ by the average power of its channel, namely $\sigma^2_{i}$. Without loss of generality, it is assumed that $\sigma^2_{i}$ is arranged in ascending order, that is $\sigma^2_{1} \leq \cdots \leq \sigma^2_{N}$, and user $m$ can be regarded as a stronger user than user $n$ if $\sigma^2_{m} \ge \sigma^2_{n}$.

\subsection{Processing Procedures}
\subsubsection{Uplink}
For uplink transmission, the optimal multiple access strategy at the transmitter is that $N$ users spread their signals across the entire bandwidth with different transmit powers. At the receiver, rather than decoding every user by treating the interference from other users as noise, the SIC technique, in which after one user is decoded, its signal is stripped away from the aggregate received signal before the next user is decoded, is applied to achieve a higher achievable rate. Specifically, first, the base station decodes the signal from the stronger user $U_i$ by treating the signals of $\{U_n\}$ as noise, where $n\in\{1,\dots ,i-1\}$. Then, the signal of $U_i$ is subtracted from the received signal to decode the signals of $\{U_n\}$. Finally, for $U_1$, it is just corrupted by the noise as the other users' signals have been successively decoded and cancelled out.

\subsubsection{Downlink}
For downlink transmission, the SIC is applied at the receiver, the superposition coding (SC) is used at the transmitter, and more power is allocated to a weaker user, namely $P_i \ge P_{i + 1}$. First, each user decodes the signals from other weaker users, i.e., $U_i$ can decode the signals of $\{U_n\}$ with $n<i$. Then, the signals of weaker users are subtracted from the received signal to decode the signal of user $U_i$, by treating the signals of $\{U_m\}$ with $m>i$ as interference. Finally, for $U_N$, it is just corrupted by the noise as the other users' signals have been successively decoded and cancelled out. Since a lower transmit power is assigned to a stronger user, the signal strength of a stronger user is not higher than that of a weaker user. Therefore, NOMA does not contradict the basic concept of SIC, in which decoding of the strongest signal should be performed first.

\subsubsection{Differences}
One of the key distinctions between the uplink and downlink transmissions is the decoding order. To be more specific, for downlink, strong users successively decode and cancel the signals of weak ones prior to decoding their own signals; whereas for uplink, the receiver successively decodes and cancels the signals of strong users prior to decoding the signals of weak ones\cite{SNOK_CST_TUTORIALS2016}.

\subsection{Capacity and Asymmetry}
Reports have said that NOMA generally outperforms OMA in the rate region with two users case\cite{DP_FWC_Cambridge0507}. In this regard, the following properties can be observed.

\subsubsection{Uplink}
Compared with NOMA, OMA is in general worse for uplink transmission except at one point, where it can achieve the same capacity bound as the former. However, at this point, the rate of the weak user is much lower than that of the strong user especially when the difference between the channel conditions of both users is large, resulting in poor fairness. To characterize the effect of the channel difference for uplink transmission, we introduce the degree of asymmetry, which is defined as the ratio between the strong and weak users' channel variances, namely ${A^u} = \frac{\sigma^2_{h_2}}{\sigma^2_{h_1}}$.

\subsubsection{Downlink}
For downlink transmission, the boundary of the NOMA rate region strictly contains the OMA rate region, leaving a gap that becomes larger as the asymmetry deepens. In those cases with severe asymmetry, however, NOMA can still provide reasonable rates for both strong and weak users. Similarly, we define the degree of asymmetry for downlink transmission as ${A^d} = \frac{\sigma^2_{h_2}}{\sigma^2_{h_1}}$.

\section{Relay-Aided NOMA Networks}
For ease of exposition, Rayleigh fading channels are considered and $N=2$ in Fig. \ref{fig_2} is assumed in this section.

\subsection{System Settings}
\subsubsection{Relay with DF protocol}
For networks with DF protocol, where the SIC is utilized at relay and user ends, % to realize multi-user detection,
the decoding order for source to relay ($S\hspace{-0.15cm}\to\hspace{-0.15cm} R$) and relay to user ($R\hspace{-0.15cm}\to\hspace{-0.15cm}U$) transmissions must be organized in the same manner. Otherwise, the achievable sum rate will decrease dramatically because of the $\min$ function, as shown in Fig. \ref{fig_3}(a). Note that, for the composite architectures, such as Figs. \ref{fig_2}(c) and (d), the decoding orders for $S\hspace{-0.15cm}\to\hspace{-0.15cm} R$ (uplink or downlink) and $R\hspace{-0.15cm}\to\hspace{-0.15cm}U$ (downlink or uplink) transmissions are different. Consequently, the channel gains for the $S\hspace{-0.15cm} \to\hspace{-0.15cm} R$ link are configured as ascending (descending) while the one for the $R\hspace{-0.15cm} \to\hspace{-0.15cm} U$ link as descending (ascending). In fact, such settings for the composite architectures are reasonable since it can reach a higher achievable sum rate and balance the throughput fairness among all multiplexed users.

\subsubsection{Relay with AF protocol}
For networks with AF protocol, where the SIC is only utilized at the user end, according to the decoding order at the user, the power allocations for $S\hspace{-0.15cm} \to\hspace{-0.15cm} R$ and $R\hspace{-0.15cm} \to\hspace{-0.15cm} U$ transmissions must be performed in the same manner. Otherwise, the system performance, e.g., the achievable sum rate and outage probability, will deteriorate significantly. A special case is that, for the uplink or downlink architecture, both $S\hspace{-0.15cm} \to\hspace{-0.15cm} R$ and $R\hspace{-0.15cm} \to\hspace{-0.15cm} U$ transmissions employ the same fixed power allocation scheme\cite{DMFYY_NOMAmultipleantennarelaying_TCOM18}. On the other hand, for the composite architecture, the channel gains for $S\hspace{-0.15cm} \to\hspace{-0.15cm} R$ and $R\hspace{-0.15cm} \to\hspace{-0.15cm} U$ links should be sorted in the same manner, namely both in ascending or descending orders. Unfortunately, such setting is impractical because it will result in a poor throughput fairness among all users. Therefore, an uplink or downlink architecture for the AF relay network could be a better choice.

\subsection{System Asymmetry}
\begin{figure}[p]
\centering
\includegraphics[width=6.0in]{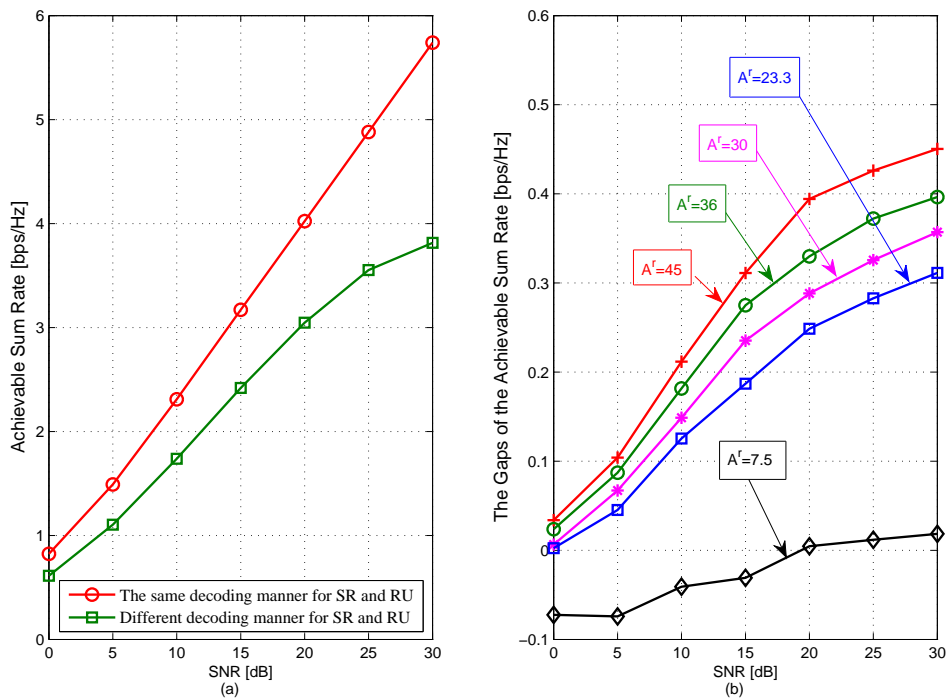}
\caption{\small Comparison between NOMA\cite{WMYYFF_NOMA_ICSI} and OMA Max-min schemes: (a) achievable sum rate with DF protocol under different decoding manners, where the channel setting is $\sigma^2 _{SR_1}=1$, $\sigma^2 _{SR_2}=10$, $\sigma^2 _{{R_1}U}=9$, and $\sigma^2 _{{R_2}U}=2$; (b) the gaps of achievable sum rate under different $A^r$, where the channel settings are: (1) $\sigma^2 _{SR_1}=1$, $\sigma^2 _{SR_2}=10$, $\sigma^2 _{{R_1}U}=9$, and $\sigma^2 _{{R_2}U}=2$; (2) $\sigma^2 _{SR_1}=1$, $\sigma^2 _{SR_2}=10$, $\sigma^2 _{{R_1}U}=9$, and $\sigma^2 _{{R_2}U}=3$; (3) $\sigma^2 _{SR_1}=1$, $\sigma^2 _{SR_2}=10$, $\sigma^2 _{{R_1}U}=7$, and $\sigma^2 _{{R_2}U}=3$; (4) $\sigma^2 _{SR_1}=1$, $\sigma^2 _{SR_2}=12$, $\sigma^2 _{{R_1}U}=9$, and $\sigma^2 _{{R_2}U}=3$; (5) $\sigma^2 _{SR_1}=4$, $\sigma^2 _{SR_2}=9$, $\sigma^2 _{{R_1}U}=10$, and $\sigma^2 _{{R_2}U}=3$.}
\label{fig_3}
\end{figure}

The degree of asymmetry for the cooperative relay system can be defined as ${A^r} = \frac{\sigma^2_{h_{SR}^s}}{\sigma^2_{h_{SR}^w}}\cdot\frac{\sigma^2_{h_{RU}^s}}{\sigma^2_{h_{RU}^w}}$, where $r$ denotes the relay networks, and $s$ and $w$ denote the strong and weak users for $S\hspace{-0.15cm} \to\hspace{-0.15cm} R$ and $R\hspace{-0.15cm} \to\hspace{-0.15cm} U$ links, respectively. Clearly, ${A^r}$ is the product of degrees of asymmetry for uplink and downlink transmissions. Once there exists a one-to-one mode, namely ${\sigma^2_{h_{SR}^s}}={\sigma^2_{h_{SR}^w}}$ or ${\sigma^2_{h_{RU}^s}}={\sigma^2_{h_{RU}^w}}$, we will have ${A^r} = \frac{\sigma^2_{h_{RU}^s}}{\sigma^2_{h_{RU}^w}}$ or  ${A^r} = \frac{\sigma^2_{h_{SR}^s}}{\sigma^2_{h_{SR}^w}}$. Fig. \ref{fig_3}(b) shows the gaps of achievable sum rate between the NOMA\cite{WMYYFF_NOMA_ICSI} and OMA Max-Min schemes under different $A^r$, where the OMA Max-Min scheme denotes that the relay $R_{\hat k}$ is selected to transmit the message from the source to the destination with $\hat k = \arg\mathop {\max }\limits_{k \in \left\{ {1,2} \right\}} \left\{ {\min \left\{ {\sigma _{{h_{SRk}}}^2,\sigma _{{h_{RkU}}}^2} \right\}} \right\}$. In general, the severer the asymmetry is, the greater the advantages of NOMA are. Therefore, when ${A^r}$ is large enough, e.g., ${A^r} > 3$, it is better to apply NOMA than OMA in practice. However, for the diamond network\cite{WMYYFF_NOMA_ICSI}, if the channel conditions satisfy $\hat k \ne 2$, the superiority of NOMA over OMA will weaken or even disappear, as the black curve in Fig. \ref{fig_3}(b) shows. This can be understood by the fact that since the source-relay-user channel selected by OMA is better than that for the symbol who is last decoded in NOMA, OMA achieves a larger achievable sum rate than NOMA. Hence, for the diamond network, we would prefer NOMA to OMA when not only ${A^r}$ is large enough, but also the source-relay-user channel selected by OMA is the same as that for the user who is last decoded in NOMA. For example, the channel conditions are set as (1)-(4) in the caption of Fig. \ref{fig_3}.

\subsection{Power Allocation}
An effective power allocation scheme designed for cooperative relay networks depends on the symbols' decoding order. Generally, more power should be assigned to the earlier decoded symbols while less power is allocated to the later decoded ones. For networks with DF protocol, it is more complex to perform power allocation than the ones with AF protocol, especially with global instantaneous channel state information (CSI)\cite{WMYYFF_NOMA_ICSI}. Therefore, finding a simple and effective power allocation scheme for DF relay networks is demanding.

\subsection{Performance Comparisons}

\subsubsection{Uplink architectures}
It has been shown that NOMA can be applied in the uplink transmission. However, since different users are located in different positions and experience different channel conditions, we have to face some ticklish problems. On one hand, it is difficult to realize signal synchronization at both transmitting and receiving ends, and on the other hand, realizing effective control on the transmit power among different users to avoid interference is very challenging in practice.

\subsubsection{Downlink architectures}
One of the most famous downlink architectures is the classical three-node relay network\cite{JI_CapacityanalysisNOMA_CL1515}, where a source, a half-duplex DF relay, and a user are considered, as shown in Fig. \ref{fig_2}(e). However, most existing NOMA schemes only put emphasis on the Rayleigh fading scenario\cite{JI_CapacityanalysisNOMA_CL1515}. For this reason, a NOMA scheme over Rician fading channel is proposed to handle this problem\cite{MJMW_NOMAdirectandrelaytransmission_CL1605}.
\par
\begin{figure}[p]
\centering
\includegraphics[width=6.0in]{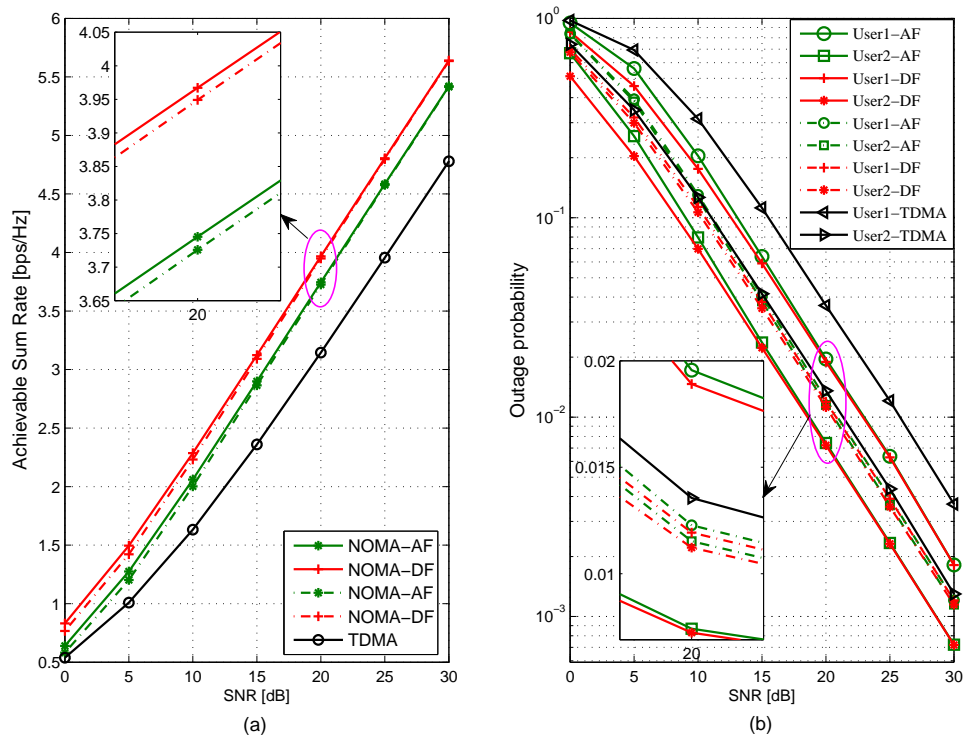}
\caption{\small Comparisons of the: (a) achievable sum rate and (b) outage probability among the NOMA-AF, NOMA-DF and TDMA schemes\cite{DMFYY_NOMAmultipleantennarelaying_TCOM18}, where the channel setting is $\sigma^2 _{SR}=8$, $\sigma^2 _{{RU_1}}=2$, and $\sigma^2 _{{RU_2}}=10$. The power allocation coefficients for user $1$ and user $2$ in NOMA are $0.6875$ and $0.3125$ (solid lines), $0.8$ and $0.2$ (dash-dot lines), respectively, while the resources allocated to user $1$ and user $2$ in TDMA are average.}
\label{fig_4}
\end{figure}

Fig. \ref{fig_2}(b) shows another downlink architecture for cooperative relay systems\cite{DMFYY_NOMAmultipleantennarelaying_TCOM18}, where a half-duplex AF or DF relay assists the communication between the source and users. Since the signal-to-interference-and-noise ratio (SINR) with DF relaying is always higher than that with AF relaying, the DF protocol is preferred over the AF one. As shown in Fig. \ref{fig_4}, the outage probability and achievable sum rate can be significantly improved when the DF protocol is applied, especially in the low signal-to-noise ratio (SNR) region. This can be understood as follows. If more power is allocated to the weak user, the system's outage behavior will be improved, while the achievable sum rate of the NOMA system will decrease since less power is allocated to the stronger user, which has a full degree of freedom in the achievable rate. Therefore, there exists a tradeoff between the outage probability and achievable sum rate.

\subsubsection{Composite architectures}
Figs. \ref{fig_2}(c) and (d) illustrate the composite architectures for DF relay systems. In general, such systems are asymmetric, where the channel gains for uplink transmission are sorted in ascending (descending) order while those for downlink transmission in descending (ascending) order\cite{MMS_NOMAXrelay_CL2017, WMYYFF_NOMA_ICSI}. Fig. \ref{fig_3}(b) shows the gaps of achievable sum rate between NOMA and OMA\cite{WMYYFF_NOMA_ICSI} with different degrees of asymmetry.
\par
Since the channel conditions for uplink and downlink transmissions are different in general, a power waste will be resulted if the transmit powers for the two links are both fixed as $P_t$. Without loss of generality, assume that the channel quality for the uplink is better than that for the downlink. In this case, we do not need to allocate full power to all symbols for the uplink, whereas for the downlink, $P_t$ must be fully allocated to all symbols. Therefore, if a central processing unit that can adjust the transmit power of the symbols according to the instantaneous CSI, the total power consumed by the system can be saved as less than $2P_t$\cite{WMYYFF_NOMA_ICSI}. On the other hand, provided that the requirement of quality of service has been satisfied, the saved power can be allocated to the weak user so as to improve its data rate, outage probability, and fairness.
\par
\begin{figure}[p]
\centering
\includegraphics[width=6.5in]{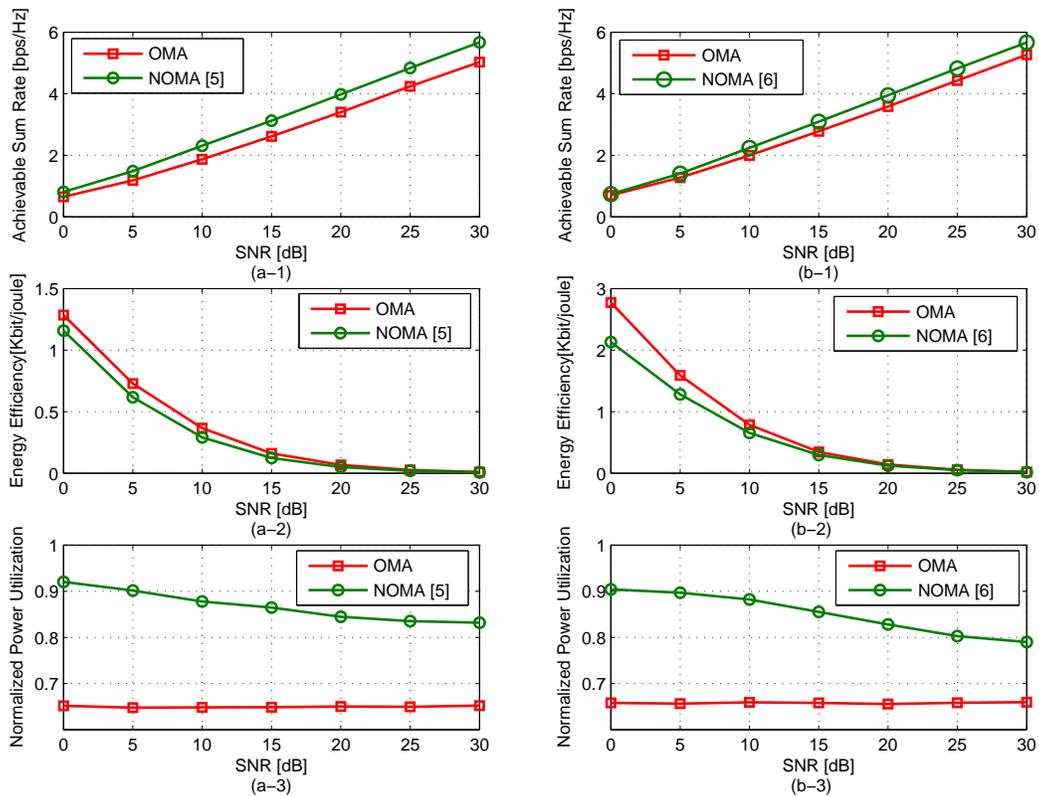}
\caption{\small Comparison in terms of achievable sum rate, energy efficiency, and normalized power utilization for: (a) the NOMA\cite{MMS_NOMAXrelay_CL2017} and OMA schemes, where the channel setting is $\sigma^2 _{S_1R}=9, \sigma^2 _{S_2R}=3$, $\sigma^2 _{{RU_1}}=2$, and $\sigma^2 _{{RU_2}}=10$; (b) the NOMA\cite{WMYYFF_NOMA_ICSI} and OMA schemes, where the channel setting is $\sigma^2 _{SR_1}=1, \sigma^2 _{SR_2}=10$, $\sigma^2 _{{R_1U}}=9$, and $\sigma^2 _{{R_2U}}=2$. The power allocation schemes applied for them are dynamic, namely adapted to the instantaneous CSIs.}
\label{fig_5}
\end{figure}

Assuming the knowledge of perfect CSIs, the achievable sum rate, energy efficiency, and normalized power utilization are chosen as performance metrics for comparison. We can see from Figs. \ref{fig_5}(a-1) and (b-1) that NOMA shows significant superiority over OMA in terms of achievable sum rate. However, oppositely, as shown in Figs. \ref{fig_5}(a-2) and (b-2), NOMA performs worse than OMA in terms of energy efficiency. This phenomenon can be explained by the fact that NOMA can fully utilize the pre-allocated power according to the channel conditions, which results in a boost in achievable sum rate, whereas for OMA, the pre-allocated power cannot be used effectively since the near-far effect exists. As shown in Figs. \ref{fig_5}(a-3) and (b-3), the real power utilization of NOMA is much larger than that of OMA, which can be up to 30\%. Although NOMA outperforms OMA in terms of achievable sum rate, the ratio between achievable sum rate and power consumption for NOMA is smaller than that of OMA, which means that the energy efficiency of the former is lower than that of the latter.

\section{Hybrid Power Allocation Strategy}
As discussed above, NOMA can be used for capacity improvement in cooperative relay systems. For the composite structures, however, design of a dynamic power allocation scheme for uplink transmission is very challenging. Fortunately, by borrowing the idea of ``hybrid-NOMA"\cite{ZMH_CooperativeNOMA_CL1513, SNOK_CST_TUTORIALS2016}, we can solve this problem with a hybrid power allocation strategy that uses both the knowledge of instantaneous CSI and statistical CSI.
\par
Assume that the network has the knowledge of the statistical CSI of the $S\hspace{-0.15cm} \to\hspace{-0.15cm} R$ link and both the statistical and instantaneous CSIs of the $R\hspace{-0.15cm} \to\hspace{-0.15cm} U$ link. The hybrid power allocation strategy can be performed in two stages. In the first stage, the power allocation scheme in\cite{MMS_NOMAXrelay_CL2017} can be employed to maximize the system's achievable sum rate with the statistical CSIs of the $S\hspace{-0.15cm} \to\hspace{-0.15cm} R$ and $R\hspace{-0.15cm} \to\hspace{-0.15cm} U$ links, by which we can obtain fixed optimal power allocation coefficients for the $S\hspace{-0.15cm} \to\hspace{-0.15cm} R$ transmission. In the second stage, by fixing the power allocation coefficients for the $S\hspace{-0.15cm} \to\hspace{-0.15cm} R$ transmission as that obtained in the first stage, we can obtain new dynamic power allocation coefficients for the $R\hspace{-0.15cm} \to\hspace{-0.15cm} U$ transmission by maximizing the system's sum rate with the instantaneous CSI of the $R\hspace{-0.15cm} \to\hspace{-0.15cm} U$ link.
\par
\begin{figure}[p]
\centering
\includegraphics[width=6.5in]{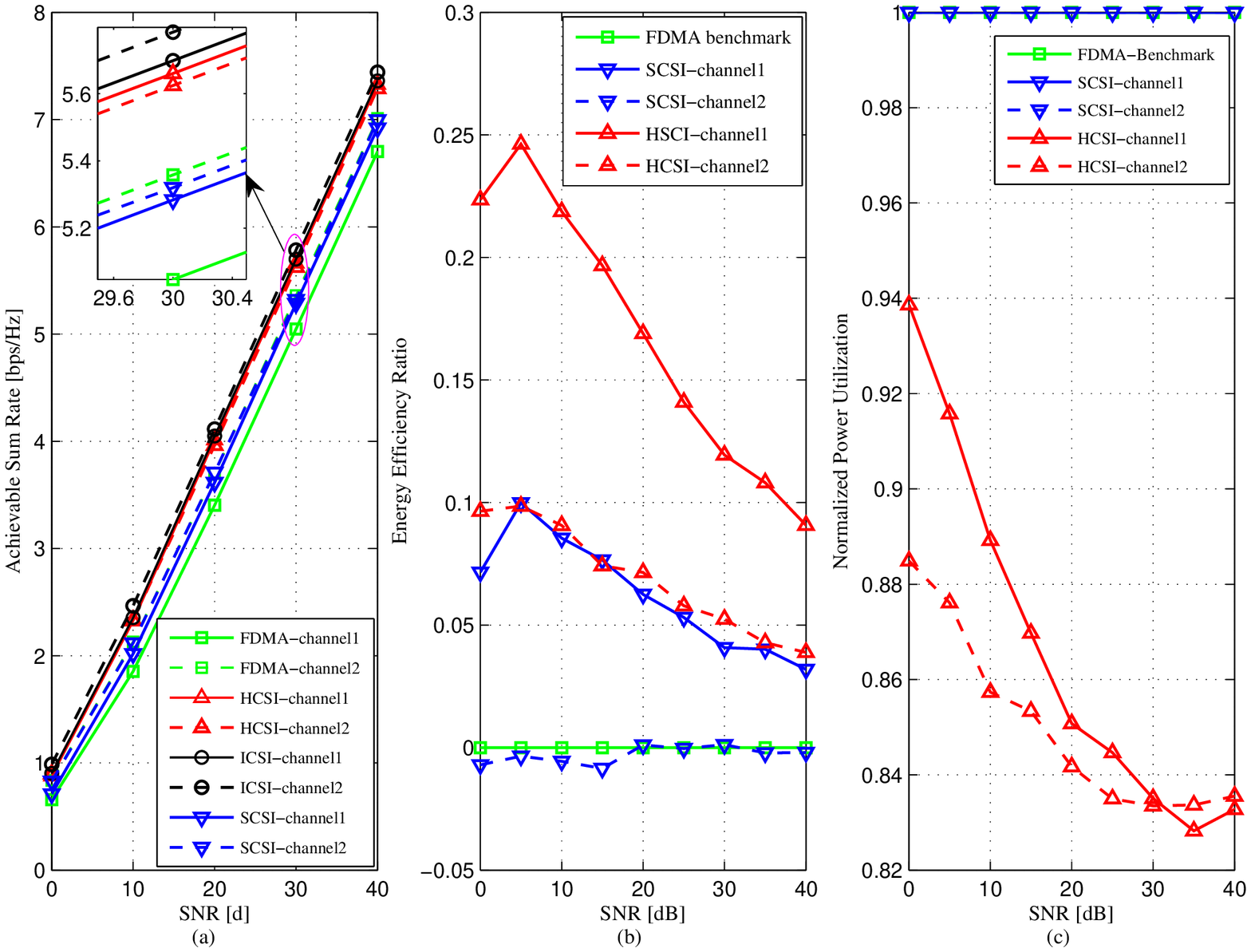}
\caption{\small Comparison in terms of: (a) achievable sum rate, (b) energy efficiency ratio, and (c) normalized power utilization among the HCSI, ICSI, SCSI, and FDMA schemes under two channel settings, namely setting $1$: $\sigma^2 _{S_1R}=9, \sigma^2 _{S_2R}=3$, $\sigma^2 _{{RU_1}}=2$, and $\sigma^2 _{{RU_2}}=10$ (solid lines) and setting $2$: $\sigma^2 _{S_1R}=9, \sigma^2 _{S_2R}=3.8$, $\sigma^2 _{{RU_1}}=5$, and $\sigma^2 _{{RU_2}}=10$ (dashed lines), where the degrees of asymmetry for the two settings are $15$ and $4.7$, respectively. The power allocation schemes applied for SCSI and FDMA are static and dependent on the statistical CSI only.}
\label{fig_6}
\end{figure}

To evaluate the performance of the hybrid power allocation strategy, we consider a specific NOMA relay system in Fig. \ref{fig_2}(c). Let ICSI and SCSI denote the power allocation schemes obtained with the knowledge of instantaneous CSI and statistical CSI for the NOMA scheme, respectively. Simulation results illustrated in Fig. \ref{fig_6} show that the hybrid power allocation strategy (marked as HCSI in the figure) can achieve almost the same achievable sum rate as the one obtained under ICSI (marked as ICSI in the figure). On the other hand, it also shows superiority over the one obtained under SCSI (marked as SCSI in the figure), even when the near-far effect is not severe. Simulation results on the energy efficiency ratio and normalized power utilization are illustrated in Figs. \ref{fig_6}(b) and (c), where FDMA is chosen as the benchmark. It is shown that with the highest spectral efficiency and lowest power consumption, the hybrid power allocation strategy performs the best in terms of energy efficiency. To sum up, considering its lower computational complexity on power scheduling, effectiveness to combat the near-far effect, and superior performance, the hybrid power allocation strategy will be a promising solution for cooperative relay systems with NOMA.

\section{Challenges, Opportunities, and Trends}

\subsection{Feedback Overhead and Computational Complexity}
For the NOMA cooperative relay schemes, the spectrum efficiency and energy efficiency can be improved significantly by applying a tailored power allocation scheme. However, such a dynamic power allocation scheme is obtained under the condition that global instantaneous CSIs can be available, which means that the signaling overhead, used for CSI and power allocation feedbacks, is large and would bring huge computational complexity especially when the number of users is very large. On the other hand, since the feedback delay will cause channel estimation error, and perfect time synchronization for the uplink is difficult to implement, obtaining perfect CSI is challenging. To circumvent this problem, new power allocation solutions, which can reduce the feedback overhead at the expense of marginal performance degradation, are demanding, such as the hybrid power allocation strategy proposed in the earlier section, and the strategy that uses limited feedback to obtain CSI\cite{PYZXR_One_Bit_Feedback_TWC2016}.

\subsection{Communication Security}
Attracted by the advantages of NOMA, recently, the secrecy issue of NOMA has been studied in\cite{YZMYL_NOMAsecurity_TWC2017}. However, how to realize secure communications with NOMA is still an open issue, especially in NOMA cooperative relay networks, since relays re-transmit a copy of the information symbols that are summed up via SC and transmitted over the same frequency band, which means that once the carrier frequency is successfully located by the eavesdropper, all users' messages may be intercepted. An effective solution is that the system should first gather CSIs and then choose an appropriate relay to perform secure communications with other relays releasing cooperative jamming. Specifically, once the appropriate relay is confirmed, the other relays should release jamming during the two phases. However, the aforementioned discussions assume that the relay works as a trusted transmitter. In fact, the scenarios with untrusted relays exist, especially in the cases that the untrusted relays with DF protocol, where all users' symbols must be decoded first before they are re-transmitted. Therefore, how to perform security communications for the NOMA cooperative relay networks in the presence of untrusted relays is an interesting issue.

\subsection{Hardware Development}
Although NOMA significantly outperforms the conventional OMA in the application to cooperative relaying, the corresponding hardware implementation associated with NOMA is more complex. For example, because of limited processing capability, it is difficult to perform multi-user detection and interference cancellation for the mobile uesr\cite{SNOK_CST_TUTORIALS2016}. Moreover, one of the reasons why NOMA outperforms OMA is that it allows more users to access the network simultaneously. Therefore, a high-performance SIC unit for the mobile receiver is a key enabler for implementation of NOMA in the future.

\subsection{NOMA Cooperative Relay Systems with mmWave}
It has been reported that the mmWave with NOMA can achieve higher spectrum and energy efficiency \cite{BLZNS_MMW_JSAC2017}. On the other hand, since the wavelength of mmWave bands is very short, effective communications generally require the transmitter and the corresponding receiver to be located in line-of-sight (LOS) range. Due to the mobility of users and surrounding obstacles, the blockage and non-line-of-sight (NLOS) channels are inevitable. In order to apply the mmWave technique to 5G systems while guaranteeing effective coverage, anti-blockage mechanisms are required through which the mmWave system is able to adaptively switch from LOS transmission mode to NLOS transmission mode. Thus, the incorporation of cooperative relay techniques into the mmWave communication system will bring many new possibilities and challenges for the deployment of mmWave cellular networks\cite{JXQMY_millimeter_wave_2016}, e.g., how to place or select the optimal relay nodes for the network will become very important.

\section{Conclusions}
In this article, we have discussed and compared the NOMA cooperative relay schemes from the aspects of basic principles, key features, criterion for system construction, and engineering feasibility. Three typical structures of the cooperative relay systems have been investigated, and simulation results have demonstrated the advantages of cooperative relaying with NOMA, especially for the composite structures. Furthermore, a hybrid power allocation strategy, which can provide lower computational complexity and reduce signaling overhead at the expense of marginal sum rate degradation, has been proposed for the NOMA-based cooperative relay networks. We have also highlighted key challenges, opportunities, and future research trends for the design of NOMA cooperative relay systems.

\section*{Biographies}
Dehuan Wan (eewan\_e@mail.scut.edu.cn) received the B.S. degree from He¡¯nan Normal University, Xinxiang, China, in 2007. He is currently working toward the Ph.D. degree with the South China University of Technology, Guangzhou, China. His recent research interests include non-orthogonal multiple access and index modulation.
\\

Miaowen Wen (eemwwen@scut.edu.cn) received the B.S. degree from Beijing Jiaotong University, Beijing, China, in 2009, and the Ph.D. degree from Peking University, Beijing, China, in 2014. From 2012 to 2013, he
was a Visiting Student Research Collaborator with Princeton University, Princeton, NJ, USA. He is currently an Associate Professor with the South China University of Technology, Guangzhou, China. He has authored a book and more than 80 papers in refereed journals and conference proceedings. His research interests include index modulation, non-orthogonal multiple access, physical layer security, and molecular communications. Dr. Wen was the recipient of the Excellent Doctoral Dissertation Award from Peking University and Best Paper Awards at IEEE ITST2012, ITSC2014, and ICNC2016. He was an Exemplary Reviewer for the IEEE Communications Letters in 2017. He currently serves as an Associate Editor of the IEEE ACCESS, and on the Editorial Board of the EURASIP Journal on Wireless Communications and Networking, the ETRI Journal, and the Physical Communication (Elsevier).
\\

Fei Ji (eefeiji@scut.edu.cn) received the B.S. degree in applied electronic technologies from Northwestern Polytechnical University, Xi¡¯an, China, and the M.S. degree in bioelectronics and Ph.D. degree in circuits and systems both from the South China University of Technology, Guangzhou, China, in 1992, 1995, and 1998, respectively. She was a Visiting Scholar with the University of Waterloo, Canada, from June 2009 to June 2010. She worked in the City University of Hong Kong as a Research Assistant from March 2001 to July 2002 and a Senior Research Associate from January 2005 to March 2005.
She is currently a Professor with the School of Electronic and Information Engineering, South China University of Technology. Her research focuses on wireless communication systems and networking.
\\

Hua Yu (yuhua@scut.edu.cn) received the B.S. degree in mathematics from Southwest University, Chongqing, China, in 1995 and the Ph.D. degree in communication and information system from South China University of Technology, Guangzhou, China, in 2004.
He was a Visiting Scholar at the School of Marine Science and Policy, University of Delaware, USA, from 2012 to 2013. He is currently a Professor at the School of Electronic and Information Engineering, South China University of Technology. He is also the Director of the National Engineering Technology Research Center for Mobile Ultrasonic Detection, Department of Underwater Communications. His research interests are in the physical layer technologies of wireless communications and underwater acoustic communications.
Dr. Yu was the Publication Chair and the Technical Program Committee Member of the 11th IEEE International Conference on Communication Systems in 2008.
\\

Fangjiong Chen (eefjchen@scut.edu.cn) received the B.S. degree in electronics and information technology from Zhejiang University, Hangzhou, China, in 1997 and the Ph.D. degree in communication and information engineering from South China University of Technology, Guangzhou, China, in 2002.
After graduation, he joined the School of Electronics and Information Engineering, South China University of Technology, where he was a Lecturer and an Associate Professor, from 2002 to 2005 and from 2005 to 2011, respectively. He is currently a full-time Professor at the School of Electronics and Information Engineering, South China University of Technology. He is also the Director of the Mobile Ultrasonic Detection National Research Center of Engineering Technology, Department of Underwater Detection and Imaging. His research focuses on signal detection and estimation, array signal processing, and wireless communication.
Dr. Chen received the National Science Fund for Outstanding Young Scientists in 2013, and was elected in the New Century Excellent Talent Program of MOE, China, in 2012.
\\
\clearpage
\clearpage
\clearpage
\clearpage
\clearpage

\begin{thebibliography}{1}
\bibitem{SNOK_CST_TUTORIALS2016}
S. M. R. Islam, N. Avazov, O. A. Dobre, and K. S. Kwak, ``Power-domain non-orthogonal multiple access (NOMA) in 5G Systems: Potentials and challenges," \emph{IEEE Communications Surveys $\&$ Tutorials}, vol. 19, no. 2, pp. 721--742, Oct. 2017.

\bibitem{SBQSSK_5G_networks_TVT_2017}
S. Chen, B. Ren, Q. Gao, S. Kang, S. Sun, and K. Niu, ``Pattern division multiple access-A novel nonorthogonal multiple access for fifth-generation radio networks," \emph{IEEE Trans. Veh. Technol.}, vol. 66, no. 4, pp. 3185--3196, Apr. 2017.

\bibitem{LBYSCZ_NOMA5G_future_CM2015}
L. Dai, B. Wang, Y. Yuan, S. Han, C.-L. I, and Z. Wang, ``Non-orthogonal multiple access for 5G: Solutions, challenges, opportunities,
and future research trends," \emph{IEEE Commun. Mag.}, vol. 53, no. 9, pp. 74--81, Sept. 2015.

\bibitem{ZYJQMCH_5G_networks_CM_2017}
Z. Ding, Y. Liu, J. Choi, Q. Sun, M. Elkashlan, C.-L. I, and H. V. Poor, ``Application of non-orthogonal multiple access in LTE and 5G networks," \emph{IEEE Commun. Mag.}, vol.~55, no.~2, pp. 185--191, Feb. 2017.


\bibitem{MMS_NOMAXrelay_CL2017}
M. F. Kader, M. B. Shahab, and S. Y. Shin, ``Exploiting non-orthogonal multiple access in cooperative relay sharing," \emph{IEEE Commun. Lett.}, vol. 21, no. 5, pp. 1159--1162, May 2017.

\bibitem{WMYYFF_NOMA_ICSI}
D. Wan, M. Wen, H. Yu, Y. Liu, F. Ji, and F. Chen, ``Non-orthogonal multiple access for dual-hop decode-and-forward relaying," in {\emph{Proc. IEEE Global Communications Conference}}, Washington DC, USA, Dec. 2016, pp. 1-6.

\bibitem{DMFYY_NOMAmultipleantennarelaying_TCOM18}
D. Wan, M. Wen, F. Ji, Y. Liu, and Y. Huang, ``Cooperative NOMA systems with partial channel state information over Nakagami-\emph{m} fading channels," \emph{IEEE Trans. Commun.}, vol. 66, no. 3, Mar. 2018.

\bibitem{ZMH_CooperativeNOMA_CL1513}
Z. Ding, P. Fan, and H. V. Poor, ``Impact of user pairing on 5G non-orthogonal multiple access downlink transmissions," \emph{IEEE Trans. Veh.
Technol.}, vol. 65, no. 8, pp. 1462--1465, Aug. 2016.

\bibitem{DP_FWC_Cambridge0507}
D.\ Tse and P.\ Viswanath, {\emph{Fundamentals of Wireless Communications}}. Cambridge University Press, 2005.

\bibitem{JI_CapacityanalysisNOMA_CL1515}
J.\ Kim and I.\ Lee, ``Capacity analysis of cooperative relaying systems using non-orthogonal multiple access," {\emph{IEEE Commun. Lett.}}, vol. 19, no. 11, pp. 1949--1952, Nov. 2015.

\bibitem{MJMW_NOMAdirectandrelaytransmission_CL1605}
R. Jiao, L. Dai, J. Zhang, R. MacKenzie, and M. Hao, ``On the performance of NOMA-based cooperative relaying systems over Rician fading channels," {\emph{IEEE Trans. Veh. Technol.}}, vol. 66, no. 12, pp. 11409--11413, Dec. 2017.

\bibitem{PYZXR_One_Bit_Feedback_TWC2016}
P. Xu, Y. Yuan, Z. Ding, X. Dai, and R. Schober, ``On the outage performance of non-orthogonal multiple access with one-bit feedback," {\emph{IEEE Trans. Wireless Commun.}}, vol. 15, no. 10, pp. 6716--6730, Oct. 2016.

\bibitem{YZMYL_NOMAsecurity_TWC2017}
Y. Liu, Z. Qin, M. Elkashlan, Y. Gao, and L. Hanzo, ``Enhancing the physical layer security of non-orthogonal multiple access in large-scale networks," \emph{IEEE Trans. Wireless Commun.}, vol. 16, no. 3, pp. 1656--1672, Mar. 2017.

%\bibitem{LQZJ_security_ICC2017}
%L. Lv, Q. Ni, Z. Ding, and J. Chen, ``Cooperative non-orthogonal relaying for security enhancement in untrusted relay networks," in \emph{Proc. IEEE International Conference on Communications (ICC)}, Paris, France, May 2017, pp. 1--6.

\bibitem{BLZNS_MMW_JSAC2017}
B. Wang, L. Dai, Z. Wang, N. Ge, and S. Zhou, ``Spectrum and energy efficient beamspace MIMO-NOMA for millimeter-wave communications using lens antenna array," \emph{IEEE J. Sel. Areas Commun.}, vol. 35, no. 10, pp. 2370--2382, Oct. 2017.

\bibitem{JXQMY_millimeter_wave_2016}
J. Zhang, X. Ge, Q. Li, M. Guizani, and Y. Zhang, ``5G millimeter-wave antenna array: Design and challenges," \emph{IEEE Wireless Commun.}, vol. 24, no. 2, pp. 106--112, Apr. 2017.

\end{thebibliography}
\end{document}